\documentclass{amsart}[12pt]
\usepackage{amsfonts,amssymb,amsthm,amsmath,graphicx}
\usepackage{latexsym}

\usepackage{epsfig}
\DeclareSymbolFont{boperators}{OT1}{cmr}{bx}{n}

\DeclareMathAccent{\karika}{\mathalpha}{boperators}{"17}
\begin{document}

           \newcommand{\be}{\begin{equation}}
           \newcommand{\en}{\end{equation}}
           \newcommand{\ee}{\end{equation}}
            \newcommand{\bee}[1]{\begin{equation}\label{#1}}
            \newcommand{\bey}{\begin{eqnarray}}
            \newcommand{\byy}[1]{\begin{eqnarray}\label{#1}}
            \newcommand{\eey}{\end{eqnarray}}
            \newcommand{\beo}{\begin{eqnarray*}}
            \newcommand{\eeo}{\end{eqnarray*}}
            \newcommand{\R}[1]{(\ref{#1})}
            \newcommand{\C}[1]{\cite{#1}}
\newcommand{\beg}{\begin{eqnarray}}
            \newcommand{\ene}{\end{eqnarray}}
            \newcommand{\mvec}[1]{\mbox{\boldmath{$#1$}}}
            \newcommand{\x}{(\!\mvec{x}, t)}
            \newcommand{\m}{\mvec{m}}
            \newcommand{\F}{{\cal F}}
            \newcommand{\ve}{\mvec{v}}
            \newcommand{\n}{\mvec{n}}
            \newcommand{\s}{\mvec{s}}
            \newcommand{\argm}{(\m ,\mvec{x}, t)}
            \newcommand{\argn}{(\n ,\mvec{x}, t)}
            \newcommand{\T}[1]{\widetilde{#1}}
            \newcommand{\U}[1]{\underline{#1}}
            \newcommand{\X}{\!\mvec{X} (\cdot)}
            \newcommand{\cd}{(\cdot)}
            \newcommand{\Q}{\mbox{\bf Q}}
            \newcommand{\p}{\partial_t}
            \newcommand{\z}{\!\mvec{z}}
            \newcommand{\bu}{\!\mvec{u}}
            \newcommand{\rr}{\!\mvec{r}}
            \newcommand{\w}{\!\mvec{w}}
            \newcommand{\g}{\!\mvec{g}}
            \newcommand{\Nabla}{\backslash \!\nabla}

\newcommand{\M}{M}
\newcommand{\al}{ a}
\newcommand{\irr}[1]{\hspace{0.2em}
  \stackrel{
   \setbox0=\hbox{\hspace{0.06em}$\displaystyle #1$\hspace{0.06em}}
   \setbox1=\hbox{\vrule width\wd0 height0.08ex depth0pt}
   \vrule width0.08ex height0.08ex depth0.475ex \box1
   \vrule width0.08ex height0.08ex depth0.475ex }{#1}
  \hspace{0.2em}}

\renewcommand{\theequation}{\thesection.\arabic{equation}}

\title{Mesoscopic theory of microcracks}
\author{C. Papenfuss (1), P. V\'an (2), W. Muschik (3)}
\markright{Mesoscopic microcracks}
\address{1: Technische Universit\"at Berlin\\
Institut f\"ur Mechanik\\
Stra\ss e des 17. Juni\\10623 Berlin}

\address{  2: Budapest University of Technology and Economics\\
Department of Chemical Physics \\
1521 Budapest, Budafoki \'ut 8}

\address{3: Technische Universit\"at Berlin\\
Institut f\"ur Theoretische Physik\\
Hardenbergstr. 36\\10623 Berlin}

\maketitle



\noindent{\bf Abstract}\\
 The mesoscopic concept is a way to deal
with complex materials with an internal
 structure within continuum mechanics. It consists of extending
the domain of the balance equations by  mesoscopic variables and
of introducing a local distribution function of these variables as
a statistical element. In our case microcracks are modelled as
penny shaped and completely characterized by their diameter and
the unit normal to the crack surface. Two examples of crack
dynamics are given as well as a possible definition of a damage
parameter. Orientational order parameters (fabric-alignment
tensors) are defined and balance like dynamic equations for them
 are derived.

\section{A model of microcracks}
Macroscopic failure of brittle materials is initiated by the propagation
of microcracks. In a simple model the microcrack is described as a flat, rotation symmetric
surface, a so called penny shaped crack. In addition we make here the following  simplifying assumptions:
\begin{enumerate}
\item The diameter of  the cracks is much smaller than the linear dimension of the continuum element.
Under this assumption the cracks can be treated as an  internal structure of the continuum
element. The cracks are assumed small enough that there is a whole distribution of crack sizes and
orientations in the volume element.
\item The  cracks are fixed to the material. Therefore their motion is coupled to the motion of
representative volume elements.
\item The cracks cannot rotate independently of the material, i. e. if they have a  nonzero rotation
velocity at all, this rotation velocity is determined by the antisymmetric part of the time derivative of the
deformation gradient of the surrounding material and it does {\em not} depend on  crack length and
orientation.   All cracks within a volume element move and rotate with the same velocity.
\item The number of cracks is fixed, there is no production of cracks, but very short cracks  are
preexisting in the virgin material.
\item The cracks cannot decrease  area, but can only
enlarge,  meaning that cracks cannot heal.
\end{enumerate}
To summarize our model  the microcrack is
characterized by a unit vector $\n$ representing the orientation of the
surface normal and by the radius  $l$ of the spherical crack surface. These parameters will be taken as the
additional  variables in the mesoscopic theory.

\section{Different approaches to damage mechanics and the mesoscopic concept}

There are two principally different possibilities to deal with
complex materials within continuum mechanics: The first way is to
introduce additional fields depending on position and time. These
fields can be any kind of internal variables \cite{MU90,MAUMU94},
or damage parameters \cite{Kac58c,MAU92}, and damage tensors
(fabric tensors) \C{Kra96b, Kan84c}.
 In damage mechanics such additional macroscopic variables have been
introduced in many different cases of materials with internal
structure like liquid crystals \cite{ER60,LES65}, polymer
solutions \cite{Hess75,MMM} and others. The other way is a so
called mesoscopic theory. The idea is to enlarge the domain of the
field quantities by an additional variable, characterizing the
internal degree of freedom connected to the internal structure of
the material. Field quantities are introduced, which  are defined
on an enlarged  space $\mathbb R^3_x\times \mathbb R_t\times \M$.
The manifold $\M$ is given by the set of values the internal
degree of freedom can take. In our case the internal degrees of
freedom are the different sizes $l$ and orientations $\n$ of
micro\-cracks. We call this way of dealing with the internal
structure of complex materials a mesoscopic theory, because it
includes more information than a macroscopic theory on $\mathbb
R^3_x\times \mathbb R_t$, but  less than a microscopic one on the
molecular level.  The domain of the mesoscopic field quantities
$\mathbb R^3_x\times \mathbb R_t\times \M$ is called mesoscopic
space.

Macroscopic quantities are calculated from mesoscopic ones as
averages over crack sizes and crack orientations. The spatial
distribution of cracks is not relevant in the sense that the
resulting macroscopic quantities are still field quantities
depending on position and time. For a treatment of the spatial
distribution of cracks and a possible coarsening process see
\cite{BreAta01a}.

In contrast to spatial averaging introduced in \cite{BazNov00}, a
nonlocal generalization of the classical Weibull-theory, the
averages in mesoscopic theory are local in space. They are
averages over different microcrack sizes and orientations in a
volume element.

Scaling properties \cite{GluSor01a} are completely out of scope of
the whole mesoscopic theory. They result from microscopic
statistical considerations. Statistical theories of fracture
describe the breakdown of material as a second order phase
transition \cite{AndAta97a,GluSor01a}, as well as a first order
phase transition \cite{SelAta91a,ZapAta97a,ZapAta99a}.

We will apply now the mesoscopic concept to damaged material with
microcracks.  The crack length can take values between a minimal
length $l_m$ of the smallest preexisting cracks and a maximal
length $l_M$, which is limited by the linear dimension of the
continuum element. The orientation of the unit vector $\n$ can be
given by an element of the unit sphere $S^2$. Therefore in the
example of microcracks the manifold $\M $ is given by $\left[  l_m
, l_M \right]  \times S^2$. The change velocities of the
mesoscopic variables $\dot{l} $ and $\mvec{u}:= \dot{\n} $ are
defined in such a way that for $\Delta t \rightarrow 0$ we have
\be l(t + \Delta t ) = l(t) + \dot{ l} \Delta t \ , \quad  \n (t +
\Delta t ) = \n(t) + \mvec{u} \Delta t \en at later times $t +
\Delta t$. The rotation velocity $\mvec{u}$ and the length change
velocity $\dot{l}$ are the components in spherical coordinates of
the crack velocity $\mvec{v}_l$ introduced in \C{Mesocrack}. In
this previous paper \C{Mesocrack} the set of additional mesoscopic
variables $\mvec{n}$ and $l$ was called directional variable.

Beyond the use of additional variables  the mesoscopic concept
introduces a statistical element, the so-called  mesoscopic
distribution function. In our case this is a distribution of crack
lengths and orientations in the continuum element at position
$\mvec{x}$ and time $t$, called here crack distribution function
(CDF). The distribution function is the probability density of
finding a crack of length $l$ and orientation $\n$ in the
continuum element.

\section{Mesoscopic balance equations}

 Now  fields as mass density, momentum density, angular
momentum density, and energy density are defined on the mesoscopic
space. For distinguishing these fields from the macroscopic ones
we add the word ``mesoscopic``. In addition we introduce the crack
number density $N$ as an extensive quantity.  The mesoscopic crack
number density $ N (l, \n, \mvec{x}, t)$
 is the number density, counting only  cracks of length $l$ and orientation
$\n$. For this crack number density there is a balance equation
too, as it is an extensive quantity. The crack number density  can
be prescribed  independently of the mass density, although the
motion of  cracks is coupled to the motion of  surrounding
material in our model. Therefore we distinguish here between the
two fields mass density $\rho$ and number density $N$, although
they have the same equation of motion and where not distinguished
in an earlier paper \C{Mesocrack}.

\subsection{Definition of the distribution function}

Due to its definition as probability density the distribution
function is the number fraction \bee{4} f(l, \n , \mvec{x} ,t)
=\frac{  N(l, \n, \mvec{x} , t)}{ N(\mbox{\boldmath $x$},t)}\quad
, \ee in volume elements, where the number density $N(\mvec{x},t)$
is non-zero.  Here $N(\mbox{\boldmath $x$},t)$ is the macroscopic
number density of cracks of any length and orientation. As the
distribution function in equation \R{4} is not well defined if
$N(\mbox{\boldmath $x$},t) =0$, we define in addition that in this
case $f(l, \n, \mvec{x} ,t)=0$. As there is no creation of cracks
in our model the distribution function will be zero for all times
in these volume elements. In all other volume elements with a
nonzero crack number it is normalized \bee{3}
\int_{l_m}^{l_M}\int_{S^2} f(l,\n ,\mvec{x},t)l^2 d^2n dl =1 \quad
. \ee

\subsection{Balance equations of mass, momentum, angular momentum,
and energy}

For the mesoscopic densities local balance equations have been
derived from the macroscopic global ones \C{PHYSICA, MCLC, BAM,
Mesocrack}. The macroscopic balance equations express the fact
that the extensive macroscopic quantities within a region  $G$ can
change due to a flux over the boundary   $\partial G$ and due to
production and supply within $G$. This results in the general form
of a global balance equation \be \frac{d}{dt} \int_G \mvec{X} d^3
x d\mvec{n}l^2dl  = \int_{\partial G} \mvec{\phi_X}(\cdot )da +
\int_G \Sigma_X (\cdot ) d^3x d\mvec{n}l^2dl  \quad .\en A
generalized Reynolds transport theorem on the mesoscopic  space,
analogous to the one in \cite{Ehrentraut_diss}, is used to
transform the time derivative, and a generalized Gauss theorem is
applied. The boundary $\partial G$ of $G$ consists of parts in
position space, in orientation space, and in the length interval.
In regular points of the continuum we get the general form of a
local mesoscopic balance equation \cite{Mesocrack} with the
abbreviation $(\cdot) = (l, \n ,\mvec{x} ,t)$: \bee{B:20}
\frac{\partial}{\partial t}\X + \nabla_x \cdot [\mvec{v}(\cdot)\X
- \mvec{S}(\cdot)] + \nabla_n \cdot [\mvec{u}(\cdot)\X -
\mvec{R}(\cdot)] + \frac{1}{l^2}\frac{\partial}{\partial l}\left(
l^2 \dot{l} (\cdot )\X - \mvec{R_l}(\cdot)\right)\nonumber \\ =
\mvec{\Sigma}(\cdot) \ee where $\mvec{R} $ and $\mvec{R_l}$ are
the non-convective fluxes over the orientational and length part
of the boundary of $G$, and $G$ is a region in $\mathbb R^3 \times
S^2 \times [ l_m , l_M ]$. The derivative with respect to the
mesoscopic variable $(l,\mvec{n})$ is represented in spherical
coordinates. In the derivation of the local balance equation it
has been supposed that there is no flux over the boundary of the
total mesoscopic space: \be \int_{- \infty}^{\infty}
\int_{S^2}\int_{l_m}^{l^M} \nabla \cdot \mvec{\phi_X}(\cdot )d^3 x
d\mvec{n}l^2dl =0 \label{flux}
  \en Otherwise such a non-zero flux term (\ref{flux}) could
be interpreted as an additional source term on the right hand side
of the equation.

Explicitly we have:\\
 {\em Balance of mass} \bee{6}
\frac{\partial}{\partial t}\varrho(\cdot)\ +\
  \nabla_{x}\cdot\{\varrho(\cdot)\mbox{\boldmath{$v$}}(\mvec{x} ,t)\}
  +\nabla_{n}\cdot\{\varrho(\cdot)\mbox{\boldmath{$u$}}(\mvec{x} ,t)\} + \frac{1}{l^2} \frac{\partial}{\partial l}
\left( l^2 \dot{l} \varrho(\cdot) \right)\ =\ 0. \ee

{\em Balance of momentum}
\bey
  \frac{\partial}{\partial t}[\varrho(\cdot)\mbox{\boldmath{$v$}}(\mvec{x} ,t)]
+ \nabla_{x}\cdot\left[\mbox{\boldmath{$v$}}(\mvec{x}
,t)\varrho(\cdot)
  \mbox{\boldmath{$v$}}(\mvec{x} ,t) - \mvec{t}^{\top}(\cdot)\right]\ +\
\nonumber\\ + \nabla_{n}\cdot\left[\mvec{u}(\mvec{x}
,t)\varrho(\cdot)
  \mbox{\boldmath{$v$}}(\mvec{x} ,t) - \mvec{ T}^{\top}(\cdot)\right]  + \frac{1}{l^2}
\frac{\partial}{\partial l}\left( l^2 \dot{l} \varrho(\cdot)
\mbox{\boldmath{$v$}}(\mvec{x} ,t) -
 \mvec{\tau}(\cdot) \right)\ \nonumber \\  =\
 \varrho(\cdot)\mbox{\boldmath{$f$}}(\cdot).\label{7}
\eey

Here $\mvec{f}(\cdot)$ is the external acceleration density, $
\mbox{\boldmath{$t$}}^{\top}(\cdot)$ the transposed mesoscopic
stress tensor, and $ \mvec{ T}^{\top}(\cdot)$ the transposed
stress tensor on orientation space (non-convective momentum flux
in orientation space), $\tau(\cdot )$ is the momentum flux vector
with respect to the crack length variable. We introduced already
the assumption that the material velocity $\mvec{v}$ and the
rotational  velocity $\mvec{u}$ are the same for cracks of all
orientations and lengths.  \vspace{.3cm}\newline

{\em Angular Momentum}\\
The balance of angular momentum has to be taken into account as an
additional equation independent from the balance of momentum,
because there is an internal angular momentum due to crack
rotations. The total angular momentum
 \be
\mbox{\boldmath{$S$}}(\mvec{x} ,t) :=
\mbox{\boldmath{$x$}}\times\mbox{\boldmath{$v$}}(\mvec{x} ,t) +
\mbox{\boldmath{$s$}}(\cdot), \ee
 is the sum of the moment of
momentum and the internal angular momentum. \bey
\frac{\partial}{\partial
t}[\varrho(\cdot)\mbox{\boldmath{$S$}}(\cdot)] +
\nabla_{x}\cdot\left[\mbox{\boldmath{$v$}}(\mvec{x}
,t)\varrho(\cdot)
  \mbox{\boldmath{$S$}}(\cdot) -
  (\mbox{\boldmath{$x$}}\times\mvec{T}(\cdot))^{\top}  -
\mvec{\Pi}^{\top}(\cdot)\right]\ + \nonumber \\
+\nabla_{n}\cdot\left[\mvec{u}(\mvec{x} ,t)\varrho(\cdot)
  \mbox{\boldmath{$S$}}(\cdot) -
  (\mbox{\boldmath{$x$}}\times\mvec{
\tau(\cdot)})^{\top} - \mvec{ \pi}^{\top}(\cdot) \right]\ +
\frac{1}{l^2} \frac{\partial}{\partial l}\left( l^2 \dot{l}
\varrho(\cdot) \mbox{\boldmath{$S$}}(\cdot) -
 \mvec{\omega}(\cdot) \right)\
=\nonumber \\
=\varrho(\cdot)\mbox{\boldmath{$x$}}\times\mbox{\boldmath{$k$}}(\cdot)\
+\ \varrho(\cdot)\mbox{\boldmath{$g$}}.\label{sb1} \eey Here
$\mvec{n} \times \mvec{g}$ is the vector of couple forces (acting
on crack orientation),
 the tensor $\mvec{\Pi}$ is the surface torque, and
$\mvec{ \pi}$ is the analogue with respect to orientation, and
$\omega$ is the analogue with respect to crack length. These
constitutive quantities appear in the non-convective fluxes in
position space, orientation space, and in the length interval,
respectively. This equation can be simplified with the assumptions
that the material velocity and the rotation velocity depend only
on position and time $\mvec{v}(\mvec{x},t)$ and
$\mvec{u}(\mvec{x},t)$.

However, the spin is only relevant, if the model allows for crack
rotations independently from the rotations of material elements,
and this is not the case in our simplified example dynamics.

 \vspace{.3cm}

 Similarly the balance of
energy can be given, which is omitted here and can be found in
\cite{Mesocrack}. In all balance equations in addition to the flux
with respect to the position variable there appear additional flux
terms with respect to the additional mesoscopic variables crack
orientation and length.

{\em Balance of crack number} \\
In our model the cracks move together with the material element.
Therefore their flux is the convective flux, having a part in
position space, a part in orientation space, and a part in the
length interval. There is no production and no supply of crack
number. Therefore we have for the crack number density $N$:
\bee{C} \frac{\partial}{\partial t}N(\cdot) +
  \nabla_{x}\cdot\{N(\cdot)\mbox{\boldmath{$v$}}(\mvec{x} ,t)\}
  +\nabla_{n}\cdot\{N(\cdot)\mbox{\boldmath{$u$}}(\mvec{x} ,t)\} + \frac{1}{l^2}
  \frac{\partial}{\partial l}
\left( l^2 \dot{l} N(\cdot) \right)\ =\ 0. \ee In a fixed volume
element this crack number density is proportional to the mass
density, and therefore these two fields where not distinguished in
an earlier paper \C{Mesocrack}.

We obtain from the mesoscopic balance of crack number density
\R{C} a balance of the CDF $f(l,\n, \mvec{x}, t)$,  by inserting
its definition \R{4}:

\begin{eqnarray}
  \frac{\partial}{\partial t} f(l,\n, \mvec{x}, t) +
  \nabla_{x}\cdot \left( \mbox{\boldmath{$v$}} (\mvec{x}, t) f(l,\n, \mvec{x}, t)\right)
  +\nonumber \\
  \nabla_{n}\cdot \left( \mbox{\boldmath{$u$}} ( \mvec{x}, t) f(l,\n, \mvec{x}, t)\right)
 + \frac{1}{l^2}\frac{\partial }{\partial l}
\left( l^2 \dot{l} f(l,\n, \mvec{x}, t) \right) \nonumber\\
  =
\frac{-f(l,\n, \mvec{x}, t)}{N(\mbox{\boldmath{$x$}},t) } \left
(\frac{\partial}{\partial t} + \mbox{\boldmath{$v$}}(\mvec{x},t)
\cdot
  \nabla_{x}\right )   N(\mbox{\boldmath{$x$}},t) \nonumber \\
  =\frac{-f(l,\n, \mvec{x}, t)}{N(\mbox{\boldmath{$x$}},t) }
\frac{d N(\mbox{\boldmath{$x$}},t) }{dt}=0 .\label{f_dgl}
\end{eqnarray}
The right hand side is equal to zero, as for the co-moving
observer the total number of cracks in a volume element does not
change in time, as in our model no cracks are created.  In our
model all cracks in a volume element move with the translational
velocity of the volume element $\mvec{v}(\mvec{x},t)$ and rotate
with the
 velocity $\mvec{u}= \nabla \times \mvec{v}(\mvec{x},t)$. Therefore the first three terms on the left hand
side can be summarized as  a co-moving  time derivative of the
distribution function (the time derivative of an observer moving
with the material elements) with the abbreviation
$\frac{d^c}{dt}$:
\begin{eqnarray}
 \frac{\partial}{\partial t} f(l,\n, \mvec{x}, t) +
 \mbox{\boldmath{$v$}} (\mvec{x}, t)  \cdot  \nabla_{x}f(l,\n, \mvec{x}, t)
 +\nonumber \\
 \mbox{\boldmath{$u$}} ( \mvec{x}, t)  \cdot  \nabla_n f(l,\n, \mvec{x}, t)
  = \frac{d^c f(l,\n, \mvec{x}, t) }{dt}\quad .
\end{eqnarray} If we assume in addition an incompressible motion
$\nabla_x\cdot \ve =0$, we end up with the equation of motion for
the CDF: \be \frac{d^c f(l,\n, \mvec{x}, t) }{dt}+
\frac{1}{l^2}\frac{\partial }{\partial l} \left( l^2 \dot{l}
f(l,\n, \mvec{x}, t)\right) =0 \ee This is not yet a closed
differential equation for the CDF as long as no expression for the
length change velocity of the crack $\dot{l}$ is given. An example
of such a closed equation will be discussed later.

Macroscopic quantities are obtained from mesoscopic ones as averages with
the CDF as probability density:
\begin{equation}
A(\mvec{x} ,t) = \int_{l_m}^{l_M}\int_{S^2} A(l, \n, \mvec{x} ,t
)f(l, \n ,\mvec{x},t) d^2 n l^2dl
\end{equation}

{\em Entropy balance}\\
Besides these mesoscopic balances the entropy balance is necessary
for introducing the second law of thermodynamics. Because the
production of  mesoscopic entropy is not necessarily positive for
each crack length and orientation, the entropy balance is only
interesting in its macroscopic form.
\begin{eqnarray}
  \frac{\partial}{\partial t}[\varrho(\mbox{\boldmath{$x$}},t)\eta(
  \mbox{\boldmath{$x$}},t)] +
  \nabla_{x}\cdot[\varrho(\mbox{\boldmath{$x$}},t)\eta(\mbox{\boldmath{$x$}},t)
  \mbox{\boldmath{$v$}}(\mbox{\boldmath{$x$}},t)
+ \mbox{\boldmath{$\phi$}}(\mbox{\boldmath{$x$}},t)] = \nonumber\\
  =\varrho(\mbox{\boldmath{$x$}},t)\sigma(\mbox{\boldmath{$x$}},t)\label{B:31}
\end{eqnarray}
($\eta$ = specific entropy density, $\mbox{\boldmath{$\phi$}}$ =
entropy flux density, $\sigma$ = entropy production density). The
second law is expressed by the {\em dissipation inequality}
\begin{equation}
  \sigma(\mbox{\boldmath{$x$}},t) \geq  0\label{B:33}
\quad .\end{equation}

The set of balance equations is not a closed system of equations,
 constitutive equations for mesoscopic quantities are needed.
The domain of the constitutive mappings  is the state space; here
a mesoscopic one. There are the possibilities that the mesoscopic
state space includes {\em only } mesoscopic quantities, or that it
includes mesoscopic {\em and} macroscopic quantities, and there
are examples where such mixed state spaces cannot be avoided
\cite{Journal}. (For instance in the case of liquid crystals the
macroscopic alignment tensor is included in a mesoscopic state
space. This is necessary to account for the orienting mean field
of surrounding ordered particles. Otherwise it is not possible to
describe the phase transition from the isotropic phase to the
ordered liquid crystalline phase.) Constitutive equations have to
be such that the second law of thermodynamics is fulfilled by any
solution of the macroscopic balance equations with the
constitutive equations inserted \C{amendment}. This requirement
restricts the possible constitutive functions.

Finally, even for the exploitation of the dissipation inequality,
which is possible only on the macroscopic level, the choice of
variables can be motivated by the mesoscopic background
\C{BLENK92, ZAMM92}. A relevance of these variables could not be
guessed from a purely macroscopic theory.

\subsection{Damage parameter and order parameters}
\subsection{Definition of a damage parameter}

The damage parameter is introduced as a macroscopic quantity
growing with progressive damage in such a way that it should be
possible to relate the change of material properties to the growth
of  the damage parameter.
We define the damage parameter as the fraction of cracks, which have reached a certain
 length $L$. The idea is that cracks of this and larger sizes considerably
decrease the strength of the material, and therefore their
fraction is a measure of the damage.
  This idea is related to the slender bar model of
  Krajcinovic \C{Kra96b}, where the damage parameter is introduced as the
number of  `broken bars` in the sample .

\be D (\mvec{x}, t)= \int_{L}^{\infty} \int_{S^2}f(l,\n
,\mvec{x},t)d^2n l^2dl \label{damage} \quad .\ee In this
definition of the damage parameter the possibility  of cracks of
any length ($l_M \rightarrow \infty $) is included. This is
consistent with many possible laws of crack growth, where the
crack does not stop growing.

More sophisticated definitions, taking into account the
orientational distribution too, are possible and will be discussed
elsewhere. Another measure of damage, which could be introduced is
the average crack length \cite{VanPa02}.

\subsection{Length order parameters}

From the mesoscopic distribution function two different kinds of
moment series can be built because of the dependence on  crack
length and on  crack orientation: We can introduce moments of the
distribution function with respect to crack length: \be
\int_{l_m}^{l_M} f(l,\mvec{n},\mvec{x},t) P_k(l) l^2 dl =:
p_k(\mvec{n},\mvec{x},t) \label{lmoment} \ee where $P_k (l) $ are
polynomials being orthogonal with respect to the measure $l^2 dl$:
\be \int_{l_m}^{l_M}P_i(l)P_j (l) l^2 dl = \delta{ij} \quad . \ee

The moments introduced in equation (\ref{lmoment})  still depend
on crack orientation. Averaging over all orientations gives
macroscopic fields, the length order parameters: \be
\pi_k(\mvec{x} ,t)=\int_{S^2} P_k(\mvec{n},\mvec{x},t)d^2n \quad
.\ee

In the following we will investigate the moments  of the
distribution function with respect to crack orientation.

\subsection{Orientational order parameters}

We can introduce the following set of alignment-fabric tensors of
successive tensorial order
\begin{equation}
{\al}^{(k)}({\mvec{x}},t) := \int_{l_m}^{l_M} \int_{S^2} f(l,
\n,\mvec{x},t) \irr{\underbrace{\n\circ ...\circ \n}_k}l^2 dld^2 n
\end{equation}

\noindent where $\irr{...}$ denotes the symmetric irreducible part
of a tensor \cite{EHMU98}.  Remarkable, that only the even order
tensors appear in the series because the microcracks are
represented by axial vectors, the unit normal to the crack
surface, i.e. $\mvec{n}$ and $- \mvec{n}$ are not distinguished.
Due to this symmetry all odd order moments vanish. The tensors
defined above are macroscopic quantities. We want to call them
alignment-fabric-tensors. Originally tensorial damage parameters
where introduced on a purely statistical ground, without a
mesoscopic foundation and where called 'fabric tensors of the
second kind' in damage mechanics (see Kanatani \cite{Kan84c} or
Krajcinovic \cite{Kra96b}). The alignment-fabric-tensors represent
the orientational distribution of microcracks, but do not take
into account their lengths. They have to be distinguished from the
scalar damage parameter which is a measure of the growth of
cracks. These alignment-fabric tensors form
 a whole set of internal variables in the sense of thermodynamics.

The alignment-fabric tensors are a measure of the deviation of the
crack orientation distribution from isotropy. They are all zero,
if all crack orientations are equally probable, and  at least some
alignment-fabric tensors are nonzero in case of anisotropic
distributions. The orientation distribution of cracks and
therefore the alignment-fabric tensors become important in the
dynamics of the crack distribution. There are usually the specimen
geometry and loading conditions rotation symmetric around an axis
$\mvec{d}$ (uniaxial conditions). It is reasonable to assume that
also the distribution of crack orientations is rotation symmetric
around the same axis $\mvec{d}$. Then, for symmetry reasons, all
alignment-fabric tensors of different orders can be expressed in
terms of scalar orientational order parameters $S^{(k)}$ and the
unit vector $\mvec{d}$ in the following way: \be {\al}^{(k)} =
S^{(k)}\underbrace{ \irr{\mvec{d}\circ ...\circ \mvec{d}}}_k
\qquad (k = 2,4,...) \quad , \en \noindent where the order
parameters $S^{(k)}$ are one in case of total alignment (the
microcracks stand parallel) and zero for randomly oriented cracks.

 \subsection{Equations of motion for the alignment-fabric tensors and for the
 damage parameter}

  In general a
  coupled set of equations of
motion for the alignment-fabric tensors of different order can be
derived  from the differential equation for the crack distribution
function by taking moments of this equation, i.e. multiplying with
the dyadic product $\irr{\underbrace{\n\circ ...\circ \n}_k}$ and
integrating over all orientations $\mvec{n}\in S^2$.
 This set of equations is analogous to the differential equations
for the alignment tensors in liquid crystal theory \C{PHYSICA} and
will be discussed elsewhere in more detail. In general the
equations for the different tensor orders are coupled.

 As in our model all cracks in a volume element have the same angular velocity,
namely that of the surrounding material, this set of equations
simplifies to a set of very special  balance type equations
without production, and without non-convective flux, which are not
coupled: \beg \frac{\partial {\al}^{(k)}}{\partial t} + \ve(
\mvec{x} ,t)\cdot\nabla{\al}^{(k)} +\frac{1}{2} \left( \nabla
\times \ve \right) \cdot  {\al}^{(k)} -\frac{1}{2}
{\al}^{(k)}\cdot \left(
\nabla \times \ve \right)=0\\
\mbox{or \ } \mvec{\karika{a}}^{(k)} =0 \ene for any tensor order
$k$. This special form arises due to the model assumption of a
fixed crack number and in addition cracks not moving and rotating
independently of the surrounding material. Therefore for an
observer co-moving with the material the orientation distribution
and  the alignment-fabric tensors do not change. These equations
are the equations of motion for the internal variables, which have
to be postulated in a purely macroscopic theory. For our
simplified crack dynamics the dynamics of  the alignment-fabric
tensors is not independent of the motion  of the material
elements. Therefore the change of the alignment tensors in time is
not relevant to be considered in our simplified model, as it is
completely determined by the motion of the surrounding material.
However, the situation is different for other, more complicated
crack dynamics. In any case, even if the dynamics of the tensorial
damage parameters is not interesting, the orientation distribution
itself is relevant, because of the dependence of the effective
stress on crack orientation (see below). This effective stress
determines the dynamics, as it appears for instance in the
Griffith-criterion for the onset of growth, and it also appears in
the expressions for the length change velocity discussed in the
examples below.

{\bf Orientation dependence of the effective stress}\\
 In an experiment with uniaxial tension  $\sigma$  applied to
 the sample (see figure \ref{Griffith}) the stress component $\sigma_n$,
 normal to the crack surface, depends on crack
 orientation.


 Let us assume that in an experiment a uniaxial tension $\sigma$ is applied along the
 $z$-direction.
 Then the stress component in the direction $\mvec{n}$, acting on a crack surface with surface unit normal $\mvec{n}$ is
           \be
           \sigma_{eff} = \sigma (\mvec{e}_z \cdot \mvec{n})^2 \quad
           \en
           where $\mvec{e}_z $ is the unit vector in $z$-direction. This dependence of
           the effective stress on crack orientation leads after averaging over all
           orientations to
           \beg
           \int_{S^2}  \sigma_{eff}f(l, \mvec{n},\mvec{x},t)  d^2n = \int_{S^2} \sigma
           (\mvec{e}_z \cdot \mvec{n})^2 f(l, \mvec{n},\mvec{x},t) d^2n \nonumber \\ =
           \left( \int_{S^2} \sigma
           \left( \mvec{n}\mvec{n} - \frac{1}{3}
           \mvec{\delta}\right) f(l, \mvec{n},\mvec{x},t)   d^2n + \frac{1}{3}
           \mvec{\delta}\int_{S^2} \sigma  f(l, \mvec{n},\mvec{x},t)   d^2n \right) :
           \mvec{e}_z\mvec{e}_z \nonumber \\ = \sigma \left( \mvec{a} + \frac{1}{3}
           \mvec{\delta} \right) : \mvec{e}_z\mvec{e}_z = \sigma \left(a_{zz} +
           \frac{1}{3}\right)
           \quad , \ene
           where $a_{zz}$ is the $zz$-component of the second order
           alignment-fabric-tensor $\mvec{a}$. This dependence of the effective stress
           on the alignment-fabric-tensor leads to a dependence of the crack dynamics
           (for instance the critical length) on the orientational order. Thus
            macroscopic equations of motion of damage parameters
depend on the orientational order characterized macroscopically by
the alignment-fabric tensors. Hence it
           would be interesting to study the dynamics of the alignment-fabric-tensors,
            too.

           \subsection{Differential equation for the damage parameter}
           Differentiating the  definition of the damage parameter equation
           (\ref{damage}) with respect to time  we get the following differential
           equation for the damage parameter:
           \beg
           \frac{d D(\mvec{x},t)}{dt} = \frac{d }{dt}\int_{L}^{l_M} \int_{S^2}
           f(l,\n,\mvec{x},t)d^2n l^2dl \nonumber \\
=\int_{L}^{l_M} \int_{S^2}\left(\frac{d }{dt}f(l,\n,\mvec{x},t)l^2
+ f(l,\n,\mvec{x},t)2 l \dot{l} \right)
d^2ndl \nonumber \\
=\int_{L}^{l_M} \int_{S^2}\left(- \frac{\partial}{\partial
l}\left( l^2f(l,\n,\mvec{x},t)\dot{l}\right)+ f(l,\n,\mvec{x},t)2
l \dot{l} \right) d^2ndl \nonumber \\
=- \left[ l^2f(l,\n,\mvec{x},t)\dot{l}\right]_L^{l_M}+
2\int_{L}^{l_M} \int_{S^2}f(l,\n,\mvec{x},t) l \dot{l} d^2ndl\ene

  The differential equation for the damage parameter depends on the crack distribution
  function  itself, and therefore also on the initial crack distribution, and it also
   depends   on the dynamical   equation for the crack length.

           \section{Examples of closed differential equations for the distribution
            function}
            Some model on the growth velocity of a single crack is needed in order to
            make a closed differential equation for the length and orientation
            distribution function out of equation (\ref{f_dgl}). Two different dynamics
            of crack extension from the literature  will be given here as examples.
            In the second example we suppose that for a given load not all cracks start
             growing, but only cracks exceeding a certain critical length $l_c$, which
             is given by the Griffith-criterion. As
    in many examples of a crack length change dynamics the cracks
    do not stop growing, but extend infinitely, in all these cases
    the maximal crack length has to be set to $l_M =\infty$.
    However, when the cracks become macroscopic their
    growth dynamics becomes more complicated (showing for instance branching)
    than our example dynamics.

           \subsection{Mott extension of Griffiths energy criterion  including a kinetic
            crack energy}
           When the cracks are growing the system has a kinetic energy due to the
           growth by virtue of the inertia of the
           material surrounding  the separating crack surfaces. This extension of the
           original Griffith energy concept (see below)  by a kinetic energy term goes
           back to Mott
           \cite{Mott48}: A kinetic energy term is added to the sum of the crack surface
           energy and the elastic deformation energy of the surrounding elastic material,
            and the crack length is such that the total energy of the system is constant.
            Two different loading conditions are especially interesting:    fixed loading
             ('dead weight') and 'fixed grips' conditions. In both experiments uniaxial
             symmetry is assumed. In the first case a constant force is applied to the
             ends of the specimen, leading to a tensile stress. In the second case  a
             fixed displacement is prescribed at the outer boundaries of the specimen.
             For these two loading conditions, requiring a constant total energy and an
              argument based on geometrical similarity, the following expressions for
               the crack length change velocity  have been derived (\cite{Lawn}p93):
           \\
           'Dead weight':
           \be
           \dot{l}  =\dot{ l}_T \left( 1 -
           \frac{l_0}{l}\right)\label{dot_ll})
           \quad ,\en
           where $\dot{ l}_T$ is the so called terminal velocity, not depending on crack
           length, but on the applied load $\sigma_{eff}$, and therefore on crack
           orientation.
            $l_0$ is the initial crack length.
           \\ 'Fixed grips':
           \be
           \dot{l}  =\dot{ l}_T \left( 1 - \frac{l_0}{l}\left( 2 - \frac{l_0}{l}\right)
            \left( \frac{1 +\alpha  \frac{l^2}{l_0^2}}{(1+\alpha )^2}
            \right)^{\frac{1}{2}}\right)
           \quad , \en
           where the parameter $\alpha$ is defined as
           \be
           \alpha = \frac{8\pi l_0^2}{A}\label{dot_l}
           \quad .\en
           It is the ratio of the initial crack area to the surface area $A$ of a cross
            section of the specimen.
           In the 'fixed grips' geometry the crack extension might stop again after a
           certain growth. This can be understood, because of the increase in compliance
           associated with crack extension in a finite specimen. This leads to a
           diminishing applied force and  decreasing tendency of the crack
           growth.

          From the mesoscopic point of view the growth laws
          equations (\ref{dot_l}) as well as (\ref{dot_ll}) are
          mesoscopic constitutive equations relating the length
          growth velocity $\dot{ l}$ to the external load in a
          material dependent manner.

           In both loading conditions discussed here the crack velocities have been
           derived for single cracks. If we apply these growth velocities in
our differential equation for the length distribution function,
eq. (\ref{f_dgl}) this means that  we neglect  interaction between
cracks. However, crack interactions can be taken into account by
more sophisticated expressions for the length change velocity.

             Inserting the length change velocities of the previous section into
             the differential equation for the crack distribution function, and
             integrating over all orientations leads to the following closed differential
              equations:
            \\ 'Dead weight':
           \beg
           \frac{df(l, \mvec{x},t)}{dt}  =\nonumber \\
            - \frac{1}{l^2} \frac{\partial }{\partial l}
           \left( l^2 f(l, \mvec{x},t)\dot{ l}_T \left( 1 - \frac{l_0}{l}\right)
            \right)\  . \ene
           The parameter $\dot{l}_T$ depends on the effective load $\sigma_{eff}$ and
           therefore on the second order alignment-fabric tensor.
           \\ 'Fixed grips':
           \beg
            \frac{df(l, \mvec{x},t)}{dt}  = \nonumber \\
            -\frac{1}{l^2} \frac{\partial }{\partial l}
            \left( l^2 f(l, \mvec{x},t)\dot{l}_T \left( 1 - \frac{l_0}{l}
            \left( 2 - \frac{l_0}{l}\right) \left( \frac{1 +\alpha
             \frac{l^2}{l_0^2}}{(1+\alpha )^2}\right) \right)^{\frac{1}{2}} \right)
                        \quad , \ene

 \subsection{Griffith-criterion for the onset of growth}\label{Griffith}
The criterion for the cracks to start growing adopted in the
example is the energy criterion introduced originally by Griffith
\C{Griffith}. According to Griffith \C{Griffith} there is a
criticality condition for the crack growth to start, and for
cracks larger than a critical length there is a velocity of crack
growth $\dot{l}$. From energetic considerations Griffith
\C{Griffith} derived a critical length of cracks with cracks
exceeding this length starting to grow. This critical length is
given by :

\be l_c = \frac{K}{\sigma_n^2}\quad ,\label{G} \ee where $K$ is a
material constant, and $\sigma_n$ is the stress applied
perpendicular to the crack surface. It is assumed that a stress
component within the crack plane does not cause crack growth. For
cracks smaller than the critical length $l_c$ the energy necessary
to create the crack surface exceeds the energy gain due to release
of stresses.

           \subsection{Rice-Griffith dynamics}

           A possible crack dynamics, taking into account the criticality condition of
           Griffith  is derived from a generalization of the Griffith energy criterion on thermodynamic grounds,
           introducing  Gibbs-potential \cite{Van98m}, which
           includes the stress normal to the crack surface and crack length as variables.
           The resulting crack evolution law has the form
           \beg
           \dot{l} = - \alpha  + \beta \sigma^2 l
           \quad
           \mbox{for}\quad  \ l \geq l_c \quad ,\\
           \dot{l} =0 \quad \mbox{for} \quad \ l < l_c
           \ene
           with material coefficients $\alpha$, and $\beta$.
 In case of a constant time rate of the applied stress,
           $\sigma =v_{\sigma}t$, it results:

 \beg
           \dot{l} = - \alpha  + \beta v_{\sigma}^2 l t^2
           \quad
           \mbox{for} \quad \ l \geq l_c \quad ,\\
           \dot{l} =0 \quad \mbox{for} \quad \ l < l_c \quad .
           \ene

            $v_{\sigma}$ is the time derivative of the applied  stress normal to the crack
           surface. The dependence of this normal stress on
           crack orientation leads to the following orientation
           dependence of the dynamics:

           \beg
\dot{l} = - \alpha  + \beta v_{\sigma\ 0}^2 l t^2 (\mvec{e}_z
\cdot \mvec{n})^4\quad \mbox{for} \quad \ l \geq l_c \quad ,\\
           \dot{l} =0 \quad \mbox{for} \ \quad l < l_c
           \quad ,\ene
           where $v_{\sigma\ 0}$ is the change velocity of the stress applied in the
           $z$-direction.

After averaging over all orientations this orientation dependence
leads to a dependence on the fourth moment $\int_{S^2} \mvec{n}
\mvec{n} \mvec{n} \mvec{n}f d^2n$ of the distribution function.

           This dynamics  also includes a criticality condition for the crack to start
           growing.

           With this model for the length change velocity we end up with the following
           differential equation for the distribution function:
           \beg
           \frac{df(l, \mvec{n}, \mvec{x},t) }{dt} = - \frac{1}{l^2}
           \frac{\partial }{\partial l} \left( l^2 \left(- \alpha
           + \beta v_{\sigma}(\mvec{n})^2 l t^2\right)\right) \nonumber \\
            \mbox{for}\quad l \geq l_c \\
           \frac{df(l, \mvec{n}, \mvec{x},t) }{dt} = 0 \quad \mbox{for}\quad  l < l_c
           \quad . \ene
           Solutions of this differential equation will are discussed in \cite{VanPa02}.

\section{Conclusions}

In the mesoscopic description we have introduced mesoscopic
fields, defined on an enlarged space including crack size and
orientation. Averages over crack sizes and orientations, i.e.
macroscopic quantities are calculated with a distribution function
$f$. The differential equation for this distribution function was
derived from the mesoscopic balance equations and crack growth law
for the single crack. Different such crack growth laws from the
literature were discussed.

Macroscopic quantities accounting for the progressive damage have
been defined as integrals calculated with the distribution
function. These are scalar damage parameters, like for instance
the average crack length, and Fabric-alignment tensors. For these
different scalar and tensorial damage parameters equations of
motion have been derived. The time evolution of Fabric-alignment
tensors will be of special importance under biaxial loading
conditions.

The equations of motion for the damage parameters can be compared
to the evolution equation in phase field models (or in
Landau-thoery of phase transitions. In phase field models an
additional wanted field, the phase field is introduced. The form
of the equation of motion, often in the form of a conservation law
is postulated \cite{PenFif90a, EasAta02a}. This phase field can be
compared to the damage parameter introduced here, and in the
non-unilateral case also to the Fabric-alignment-tensor. The
equation of motion for the damage parameter is of the same type.
It is a special form of a balance equation, here with a zero flux
term, because spatial inhomogeneities were not taken into account.
However, this form of equation of motion has not been postulated
here, but derived from mesoscopic considerations, i.e. mesoscopic
balance equations.

\section*{Acknowledgements}
We thank the DAAD for sponsoring the cooperation between both the
Departments of Physics and Chemical Physics  of the Technical
University of Berlin and the Budapest University of Technology and
Economics. Financial support by the VISHAY Company, 95100 Selb,
Germany, is gratefully acknowledged.


\end{document}